\documentclass[a4wide]{article}
\usepackage{epsfig,wrapfig,rotating,multirow,hyperref}

\def\Geant{{\sc Geant-4}}

\begin{document}

\title{Signatures of long-lived colored sparticles}


\author{Rasmus Mackeprang, (CERN, CH-1211 Geneva, Switzerland)}

\maketitle
\begin{abstract}
  In this paper an updated \Geant{} simulation toolkit is presented
  describing the interactions in matter of heavy hadrons containing
  long-lived squarks and gluinos. Generic signatures are derived that
  are applicable at the Large Hadron Collider, and the problem of
  theoretical uncertainties arising from model dependence is
  addressed. For a more detailed description of the assumptions and
  techniques used, the reader is referred to
  ref. \cite{Mackeprang:2009ad}.
\end{abstract}

\section{Introduction}

Stable massive particles\footnote{Here taken to mean that the decay
  length is larger than a typical collider physics experiment.} (SMPs)
are a feature of many models of physics beyond the Standard Model
\cite{Fairbairn:2006gg,Raklev:2009mg}. Exclusively weakly interacting
SMPs may be searched for utilising the induced missing transverse
energy in events where they are created. Electromagnetically charged
SMPs (e.g. sleptons) may be recognized by their ionisation properties
as they traverse a detector as well as (depending on their mass) their
time of flight. Particles carrying colour charge on the other hand
exhibit a phenomenology of somewhat larger complexity. Squarks and
gluinos for instance will form hadrons (hereafter termed
\emph{R-hadrons} in accordance with litterature) that will then
undergo not only electromagnetic but also hadronic processes in the
traversed matter. This work describes an implementation of a
\Geant\cite{Agostinelli:2002hh} toolkit describing scattering of
R-hadrons in matter. The toolkit is described in detail in
refs. \cite{Mackeprang:2009ad,webpage} and is based on previous works
\cite{Kraan:2004tz,Mackeprang:2006gx}.

\section{Scattering of heavy hadrons}

When describing the scattering of the heavy hadrons formed by
long-lived squarks or gluinos as they traverse a detector, two
questions become important: What are the stable hadronic states, and
what are their scattering properties? Answers to both questions, are
of course highly model dependent. Previous approaches
\cite{Mackeprang:2006gx,Kraan:2004tz} have employed liberal
assumptions with respect to the possible stable states allowing for
all combinations of $u$ and $d$ quarks to combine with the heavy
sparticle to form hadrons. For this work, a number of different mass
calculations have been taken into account
\cite{Chanowitz:1983ci,Farrar:1984gk,Buccella:1985cs,Foster:1998wu,Gates:1999ei}. From
these calculations a consistent set of masses can be chosen as shown
in table \ref{tab:masses}.
\begin{table}[htbp]\centering\begin{tabular}{ccc}
    \hline
    Heavy Parton & States & Mass (GeV) \\
    \hline
    Squark & ${\tilde{q}\bar{u}},{\tilde{q}\bar{d}}$ &   $m_{\tilde{q}}+0.3$  \\
    &  $\tilde{q}{ud}$  & $m_{\tilde{q}}+0.7$   \\
    \hline
    Gluino & ${\tilde{g}q\bar{q}}$,${\tilde{g}u\bar{d},\tilde{g}d\bar{u}}$, ${\tilde{g}g}$ & $m_{\tilde{g}}+0.7$  \\
     & ${\tilde{g}uds}$  & $m_{\tilde{g}}+0.7$  \\
    \hline
  \end{tabular}
  \caption{Assumed stable hadrons formed from squarks ($\tilde{q}$) and gluinos ($\tilde{g}$), together with mass estimates
    used in this work. The neutral gluino states containing a mixture $u\bar{u}$ and $d\bar{d}$ pairs
    are generically denoted by ${\tilde{g}q\bar{q}}$.}\label{tab:masses}
\end{table}

Of particular interest in table \ref{tab:masses} is the observation
that especially the baryonic states are constrained. For instance
there can be no $\tilde{t}uu^{++}$ or $\tilde{g}uuu^{++}$ states.

Turning to the scattering of the particles themselves, the toolkit
used was based on that presented in \cite{Mackeprang:2006gx}. The
software framework and the treatment of nuclear effects such as
nuclear evaporation and creation of \emph{black track} particles was
left unchanged. The innovations were the implementation of a theory
driven scattering cross section and a different approach to the
transmutation of the R-hadrons themselves.

\begin{figure}[!htbp]
  \includegraphics[width=0.4\textwidth]{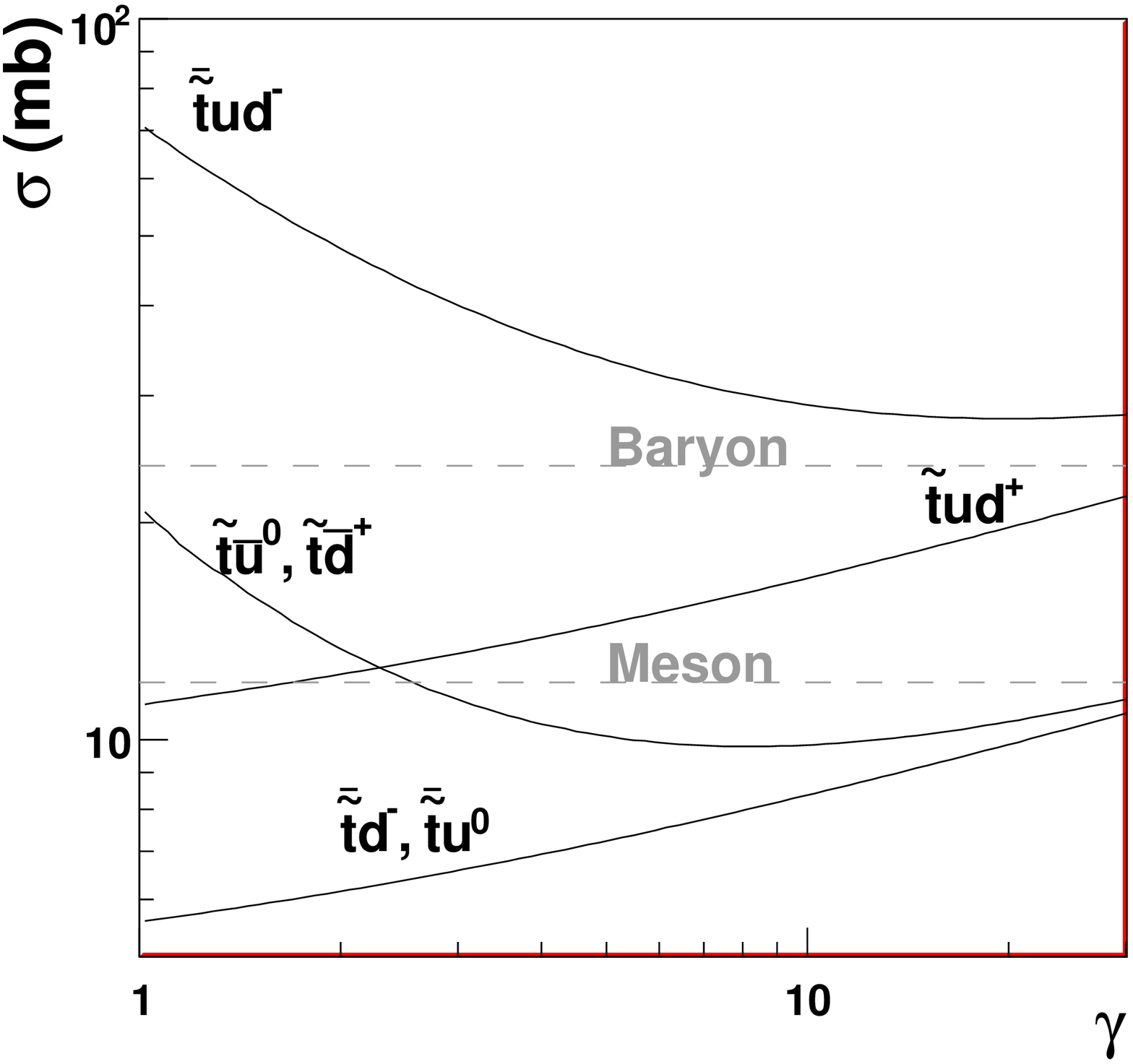}\ \hfill
  \includegraphics[width=0.4\textwidth]{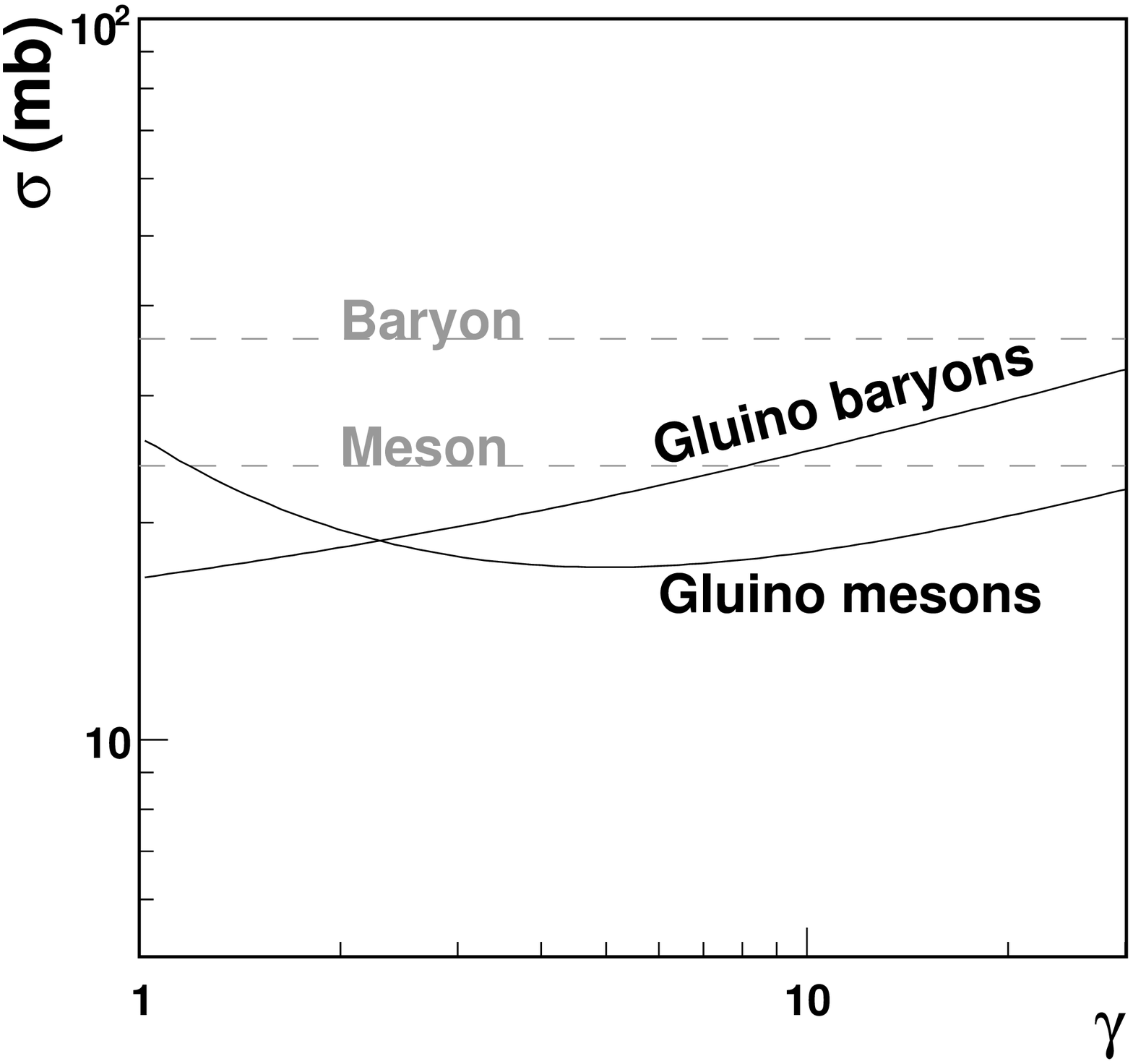}
  \caption{Cross section per nucleon for the interaction of stop-based
    and gluino-based $R$-hadrons. The predictions are shown for the
    Regge model (solid lines) and the generic model (dashed lines).}
  \label{fig_cx}
\end{figure}

The cross section calculation used stems from the model presented in
ref. \cite{deBoer:2007ii} (the \emph{Regge model}). It is depicted in
figure \ref{fig_cx} together with the predictions from the previous
model (refs. \cite{Kraan:2004tz,Mackeprang:2006gx}, hereafter referred
to as the \emph{generic model}.). For the specific formulae the reader
is referred to ref. \cite{Mackeprang:2009ad}.
Final states in the individual hadronic interaction are chosen
randomly among the kinematically allowed processes with the constraint
that meson states should turn into baryon states with a 10\%{}
probability. Once an $R$-hadron becomes a baryon it cannot turn back
into a meson by subsequent hadronic collisions as was first noted in
\cite{Kraan:2004tz}.

Another process that must be taken into account is the possibility
that neutral squark mesoninos may oscillate into their
anti-particle.
As mixing of such states would be highly model dependent we allow only
for the cases of no mixing and maximal mixing, where there is a 50\%{}
probability for any netral squark mesonino to turn into its
anti-particle. These choices correspond to oscillation lengths of zero
and infinity, respectively.

\section{Signatures deriving from hadronic interactions}

The heavy sparticle is typically orders of magnitude heavier than the
light quark system (LQS) of the R-hadron. As the scale (i.e. geometric
cross section) of a wave function is inversely proportional to the
square of the mass \cite{Kraan:2004tz}, any hadronic interaction is
assumed to take place between a nucleon and the LQS of the
R-hadron. The kinetic energy of the LQS system scaling with
$\frac{M_{LQS}}{M_{Total}}E_{kin,Total}$ therefore represents a measure of the
available energy in any collision.
\begin{figure}[!htbp]
  \epsfig{width=0.395\textwidth,file=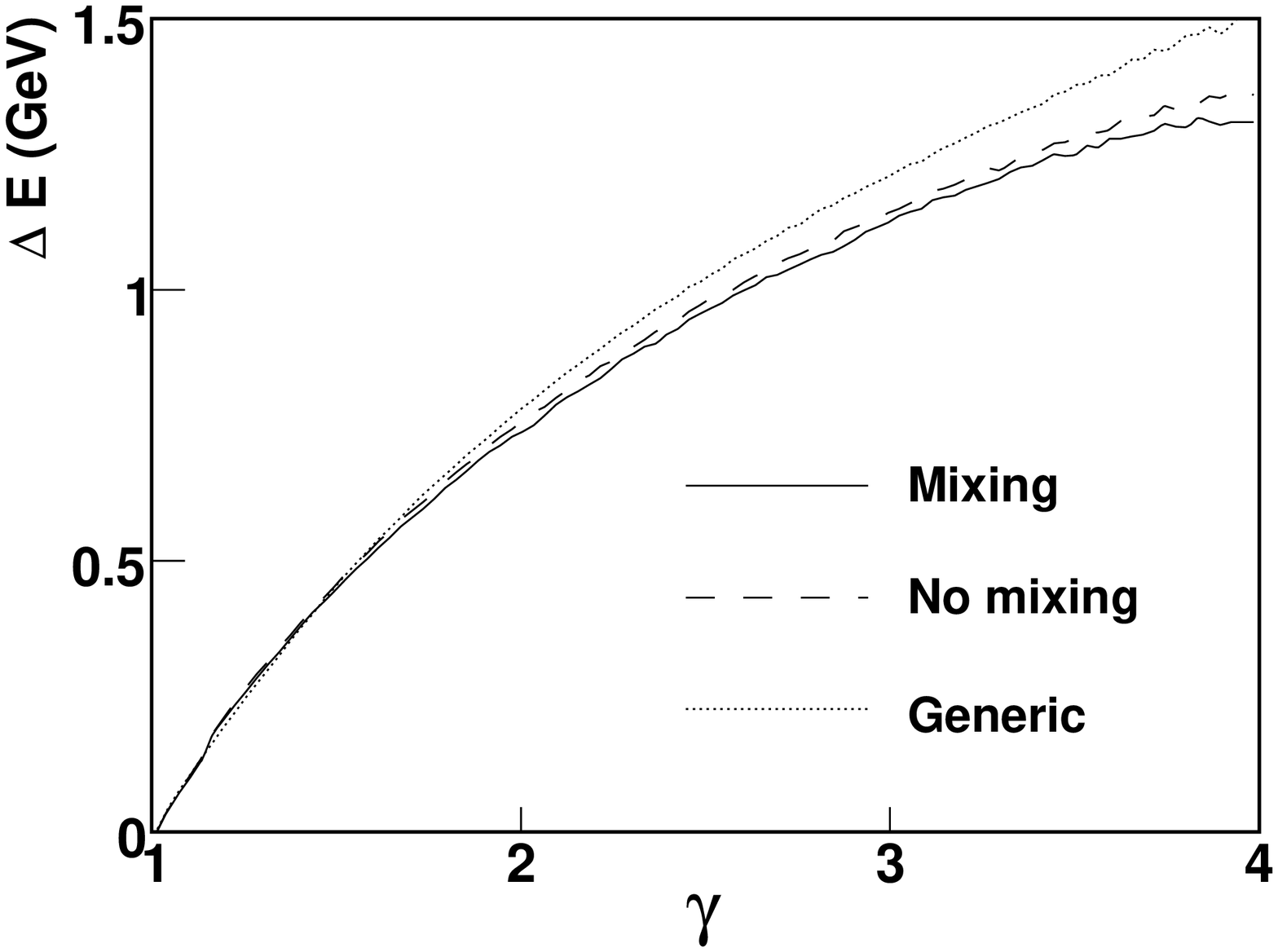}
  \epsfig{width=0.595\textwidth,file=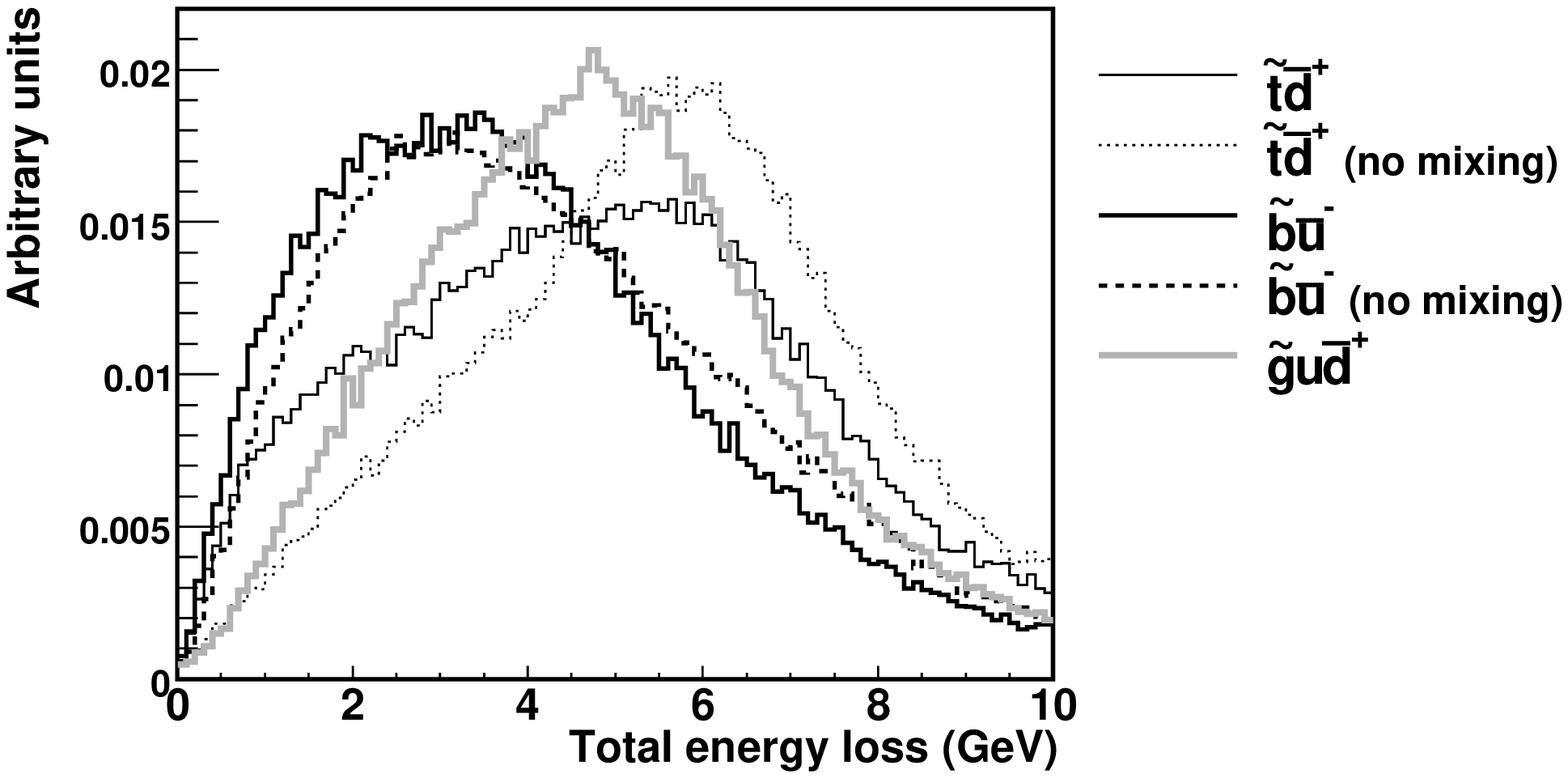}
  \caption{Left: Energy loss per hadronic interaction for stop hadrons in
    iron is shown for the Regge model with and without mixing as well
    as the generic model.
    Right: Total energy loss for R-hadrons traversing 2 m of iron
  }\label{fig:eloss}    
\end{figure}
This is consistent with the observed energy loss per hadronic
interaction shown in figure \ref{fig:eloss} (left). Generating gluino
pair-production events with Pythia\cite{Sjostrand:2006za} for
\emph{pp} collisions at a centre-of-mass energy of 14 TeV, kinematic
distributions were obtained for R-hadrons. R-hadrons were
subsequentially simulated traversing 2 m of iron to estimate their
behaviour in an LHC experiment \cite{Wigmans:2000vf}. The results
showed that the different types of R-hadrons would undergo $\sim$3-8
hadronic interactions and deposit energy as shown in figure
\ref{fig:eloss} (right). These observations imply that R-hadrons at
the LHC experiments \emph{will} interact hadronically but that they
will not through these interactions give rise to anything that might
be interpreted as hard jets.

Using similarly generated kinematics as for the energy loss
observations, the distributions of R-hadrons on the different flavours
in table \ref{tab:masses} were studied as a function of penetration
depth in iron.
\begin{figure}[!htbp]
\epsfig{file=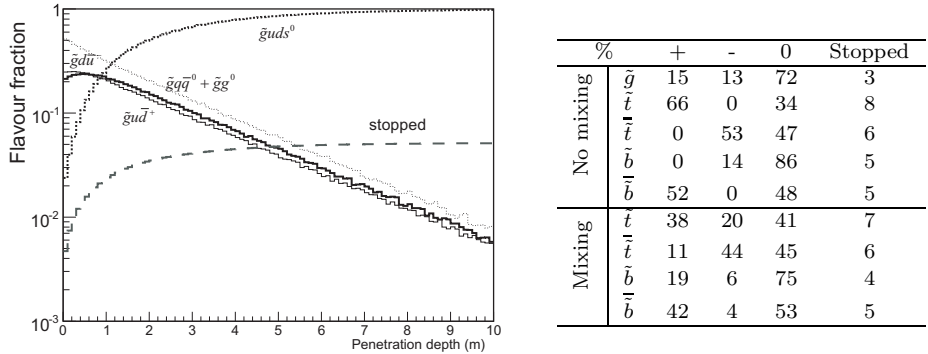,width=0.55\textwidth}\qquad
{\footnotesize
  \begin{tabular}[b]{l|lcccc}
    \hline
    \multicolumn{2}{c}{\%}  & + & - & 0 & Stopped\\
    \hline
    \multirow{5}{*}{\rotatebox{90}{\mbox{No mixing}}}
    & $\tilde{g}$ & 15 & 13 & 72 & 3\\
    & $\tilde{t}$ & 66 & 0 & 34 & 8\\
    & $\overline{\tilde{t}}$ & 0& 53 & 47 & 6\\
    & $\tilde{b}$ & 0  & 14 & 86 & 5\\
    & $\overline{\tilde{b}}$ & 52 & 0 & 48 & 5\\
    \hline
    \multirow{4}{*}{\rotatebox{90}{\mbox{Mixing}}}
    & $\tilde{t}$ & 38 & 20 & 41 &7 \\
    & $\overline{\tilde{t}}$ & 11 & 44  & 45 & 6 \\
    & $\tilde{b}$ & 19 &6 &75 &4 \\
    & $\overline{\tilde{b}}$ & 42 &4 &53 &5 \\
    \hline
    \multicolumn{2}{c}{}\\
    \multicolumn{2}{c}{}\\
  \end{tabular}
}
\caption{Left: Flavour distribution of 300 GeV gluino R-hadrons as a
  function of penetration depth in iron. Right: Percentages of
  different charges and stopping fractions at a penetration depth of 2
  m in iron for 300 GeV R-hadrons}
  \label{fig:flav}
\end{figure}
As is shown in figure \ref{fig:flav} (left), gluino R-hadrons have a
$\sim$50\%{} probability of being baryons after traversing 2 m of
iron. Of the remaining R-mesons, roughly half will be electrically
neutral. The stopped fraction of the R-hadrons is on the order of a
few percent. Comparing these numbers to the stop and sbottom cases,
one needs to consider both squarks and anti-squarks for the minimal
and maximal mixing scenarios. This is done in the table in figure
\ref{fig:flav} (right) where the gluino numbers are also included for
comparison. As can be seen, the fraction of stopped R-hadrons is in
all cases 3-8\%. Gluino and sbottom hadrons in all scenarios are found
to be predominantly neutral while the stop is found to have a charged
fraction of roughly 2/3. These observations are not greatly diluted
when mixing is enabled.

It turns out that there is little impact on these conclusions by
varying the meson/baryon mass splittings in the model. Varying the
mass splitting by 1 GeV moves the predictions less than one
percent. Varying the hadronic cross section by a factor of 2 produces
an effect of O(10\%) in the predictions. The scattering cross section
thus remains a key ingredient in the estimation of the theoretical
uncertainties for the scattering of heavy hadrons containing coloured
sparticles. Should it turn out that more baryonic states of squark or
gluino hadrons exists, the prediction of this model would of course
also be altered.

\section{Acknowledgements}
The author wishes to acknowledge his fruitful collaboration with prof. D.A. Milstead, Stockholm University.

\bibliographystyle{aipproc}   

\end{document}